\documentclass{mn2e}
\usepackage{psfig}

\def\ltsima{$\; \buildrel < \over \sim \;$}
\def\lsim{\lower.5ex\hbox{\ltsima}}
\def\gtsima{$\; \buildrel > \over \sim \;$}
\def\gsim{\lower.5ex\hbox{\gtsima}}

\begin{document}

\title[Jet structure of GRBs]{Afterglow lightcurves, 
viewing angle and the jet structure of $\gamma$-ray bursts}

\author[Rossi, Lazzati \& Rees]
{Elena Rossi, Davide Lazzati \& Martin J. Rees \\
Institute of Astronomy, University of Cambridge, Madingley Road,
Cambridge CB3 0HA, England \\ 
{\tt e-mail: emr,lazzati,mjr@ast.cam.ac.uk}}

\maketitle

\begin{abstract}
Gamma ray bursts are often modelled as jet-like outflows directed
towards the observer; the cone angle of the jet is then commonly
inferred from the time at which there is a steepening in the power-law
decay of the afterglow.  We consider an alternative model in which the
jet has a beam pattern where the luminosity per unit solid angle (and
perhaps also the initial Lorentz factor) decreases smoothly away from
the axis, rather than having a well-defined cone angle within which
the flow is uniform.  We show that the break in the afterglow light
curve then occurs at a time that depends on the viewing angle. Instead
of implying a range of intrinsically different jets -- some very
narrow, and others with similar power spread over a wider cone -- the
data on afterglow breaks could be consistent with a standardized jet,
viewed from different angles. We discuss the implication of this model
for the luminosity function.
\end{abstract}


\section{Introduction}
There are strong reasons for suspecting that the emitting plasma of
$\gamma$-ray bursts (GRBs) is geometrically beamed in a cone.  The
energy requirements can then be reduced below the exorbitant levels
that isotropic emission would imply (Kulkarni et al. 1999) and in most
models for the long bursts it is in any case a natural expectation --
borne out by simulations (MacFadyen \& Woosley 1999; MacFadyen,
Woosley \& Heger 2001) -- that the relativistic outflow from a central
engine should be collimated along a channel that opens up along the
rotation axis of a massive star.  Although the relativistic MHD that
gives rise to jets is uncertain, and surely very complicated, most
discussions of the radiation from gamma ray bursts (and their
afterglows) has postulated a jet with a well-defined angle (Meszaros
\& Rees 1997), though this angle may differ for different bursts (see,
however, Meszaros, Rees \& Wijers 1998; Salmonson 2001). We
discuss here an alternative model where the jet, rather than having a
uniform profile out to some definite cone angle, has a ``beam
pattern'' where the power per unit solid angle (and perhaps also the
initial Lorentz factor) is maximal along the axis, but drops off
gradually away from the axis.  This would be expected if there is
mixing and entrainment from the borders of the funnel.  We discuss the
expected time-dependence of the afterglow if it is triggered by a jet
with this more general profile. We conclude that the time of the
observed break (which, for a uniform jet viewed along its axis,
depends on the cone angle; see, e.g., Rhoads 1997) instead depends on
the angle between the line of sight and the symmetry axis: such a jet
viewed nearly head-on simulates a narrow uniform jet, whereas the
afterglow from the same jet viewed more obliquely would simulate a
wider uniform jet. GRB have been proposed recently to be
explosions releasing a standard power that can be either injected in
very different jet opening angles or distributed within the jet in
some universal emission diagram, (Postnov et al. 2001). While Frail et
al. (2001, hereafter F01), support with their data the first
interpretation, we show that their observational results could instead
be attributed to a more standard set of objects viewed at different
angles to their symmetry axis.

\section{Dynamics and break time}

\noindent  
 We suppose that all long GRBs have jets with a standard opening angle
$\theta_{j}$, total kinetic energy and beam profile for
$0<\theta<\theta_{j}$.  We consider a relativistic outflow where both
the bulk Lorentz factor and the energy per unit solid angle depend as
power laws\footnote{ We concentrate here on an power index $-2$
motivated by F01 correlation. This is not, however, a necessary
ingredient of the model as discussed in \S6 (see also Fig. 3 and 4)}
on the angular distance from the center $\theta$
\begin{equation}
\epsilon=\left\{\begin{array}{ll}
             \epsilon_{c} & \;\;\;0 \leq \theta \leq \theta_{c}\\
             \epsilon_{c}\big(\frac {\theta}{\theta_{c}})^{-2} & \;\;\;\theta_{c} \leq \theta \leq \theta_{j}
	\end{array}\right.
\label{eq:top}
\end{equation}
and 
\begin{equation}
\Gamma=\left\{\begin{array}{lll}
       \Gamma_{c}& &\;\;\;0 \leq \theta \leq \theta_{c}\\
       \Gamma_{c}\big(\frac {\theta}{\theta_{c}})^{-\alpha_{\Gamma}},&\alpha_{\Gamma}>0
& \;\;\; \theta_{c} \leq \theta \leq \theta{j},
     \end{array}\right.
\label{eq:gammadis}
\end{equation}
where $\theta_c$ is introduced just for formal reasons to avoid a
divergence at $\theta=0$, but can be taken to be smaller than any
other angle of interest. A lower limit to this angle is
$\theta_c>1/\Gamma_{\max}\sim10^{-3}$ degrees, where $\Gamma_{\max}\sim
10^5$ is the maximum value to which the fireball can be accelerated
(Piran 1999).  The power law index of $\Gamma$, $\alpha_{\Gamma}$, is
not important for the dynamics of the fireball and the computation of
the light curve as long as
$\Gamma(t=0,\theta)\equiv\Gamma_{0}(\theta)>\theta^{-1}$ and
$\Gamma_{0}(\theta) \gg 1,\,\forall \theta$.  Nevertheless, it plays a
role when we want to calculate the fraction of GRBs-afterglow without
prompt $\gamma$-ray emission, as we discuss in \S 6.

Consider an observer at an angle $\theta_{o}<\theta_{j}$ with respect
to the axis of the jet. He measures an isotropic equivalent energy
$E_{iso}=4\pi\epsilon(\theta_{o})$ from the $\gamma$-ray fluence.  If
also the afterglow emission is dominated by the component pointing the
earth, he will infer $\theta_{o}$ as the half-opening angle of the jet
by means of the break time in the light curve, $t_{b}$ (Sari, Piran \&
Halpern 1999) $\theta_{j}\propto t_{b}^{3/8}\, (E_{iso}/n)^{-1/8}$,
where $n$ is the external medium density.  Since the total energy
inferred from all viewing angles is $E_{tot}\simeq
2\pi\,\theta^{2}\epsilon=$const, the observer will derive the same
conclusions obtained by F01 and Panaitescu \& Kumar (2001, hereafter
PK01) of GRBs as fireballs with the same total kinetic energy but
very different jet apertures.

To evaluate how the contributions of the other components add to the
light curve from the zone with $ \theta \sim \theta_{o}$ we calculate
when their beamed emission include the observer direction, (when
$\theta-\theta_{o}<\frac{1}{\Gamma}$) and which energy per unit solid
angle they have at that time compared to
$\epsilon(\theta_{o})\equiv\epsilon_{o}$.  We show that under the
assumptions of Eq.~\ref{eq:top} the afterglow light curve is indeed
dominated by the fireball element along the line of sight: neither the
``core'' of the jet with $\theta\ll\theta_o$ nor the regions with
$\theta\gg\theta_o$ make substantial contributions.

For an effective though simplified discussion we consider only three
components of the jet corresponding to
$\theta=\theta_{1}\ll\theta_{o}$, $\theta=\theta_{o}$ and
$\theta=\theta_{3}\gg\theta_{o}$.  We call them ``cone $i$'', where
$i=1,o,3$, respectively.  This approach is justified by the fact that
only the very inner parts of the jet with $\epsilon \gg \epsilon_{o}$
and the much wider ($\pi\theta^{2} \gg \pi\theta_{o}^{2}$) outer parts
could contribute and substantially modify the light curve of cone $o$.
We approximate cone $1$ as a relativistic source moving at an angle
$\theta_{o}$ with respect the observer, while the observer is
approximately considered on the symmetry axis for cones $o$ and $3$.
The cones, that are not causally connected ($\Gamma_0>\theta^{-1}$ ),
evolve independently and adiabatically in a constant density medium
and spread sideways when $\Gamma_i$ drops below $\theta_i^{-1}$
(Rhoads 1997).  We consider relativistic lateral expansion of the
cones so that their geometrical angles $\theta_{i}$ grow as
$\theta_{i}\sim\theta_{i}(t=0)+\frac{1}{\Gamma}_i.$ Since the dynamics of the
fireball components changes at $t_b$, also $\Gamma$ decreases with
time differently before and after the break time.  For $t < t_{b}$
\begin{equation}
\Gamma=\left\{\begin{array}{ll}
     \Gamma_{0}(\frac{t}{t_{d}})^{-3/2} & \textrm{for $\theta_1$}\\
     \Gamma_{0}(\frac{t}{t_{d}})^{-3/8} & \textrm{for $\theta_o$ and $\theta_3 $,} \\
             \end{array}\right.
\label{eq:gamma}
\end{equation}
where $t_{d}$ is the deceleration time at which $\Gamma=\Gamma_{0}/2$
and the jet begins to decelerate significantly.  From
Eq.~\ref{eq:gamma} to Eq.~\ref{eq:tb} we drop the subscript $i$ for an
easier reading.  For $t > t_{b}$
\begin{equation}
\Gamma=\theta^{-1}\left(\frac{t}{t_{b}}\right)^{-1/2}.
\label{eq:gammatb}
\end{equation}
\noindent
This on-axis calculation is valid for all $\theta_i$, because
$\Gamma_1\sim1/\theta_o$ soon after the break time and the emitting
plasma enters the line of sight: from this moment we can consider the
observer to be on the cone axis.  The lateral expansion starts at a
time
\begin{equation}
t_{b}=\left\{\begin{array}{ll}
     \theta^{2/3}\, \Gamma_{0}^{2/3}\,t_{d} & \textrm{for $\theta_1$}\\
     \theta^{8/3}\, \Gamma_{0}^{8/3}\,t_{d} & \textrm{for $\theta_o$ and $\theta_{3}$,} \\

             \end{array}\right.
\label{eq:tb1}
\end{equation}
where the deceleration time, $t_{d}$, depends on
\begin{equation}
t_{d}\propto\left\{\begin{array}{ll}
     \epsilon^{1/3}\, \Gamma_{0}^{-2/3} (1-\beta\cos\theta_{o})& \textrm{for $\theta_1$}\\
     \epsilon^{1/3}\, \Gamma_{0}^{-8/3} & \textrm{for $\theta_o$ and $\theta_{3}$,} \\
     
             \end{array}\right.
\label{eq:td}
\end{equation}
and consequently 
\begin{equation}
R_{b}\propto\theta^{2/3}\epsilon^{1/3}.
\label{eq:rb2}
\end{equation}
Therefore from Eq.~\ref{eq:top} it follows that  $R_{b}$=constant and
\begin{equation}
t_{b}\propto\left\{\begin{array}{ll}
      \theta^{2/3}\,\,\epsilon^{1/3}\,\theta_{o}^{2}\propto \theta_{o}^{2}& \textrm{for $\theta_1$}\\
     \theta^{8/3}\,\,\epsilon^{1/3}\propto\theta^{2}& \textrm{for $\theta_o$ and $\theta_{3}$.}\\
             \end{array}\right.
\label{eq:tb}
\end{equation}
Cones $o$ and $3$ are hollow but they spread only outwards because the
inner components which have already expanded (for cone $3$) or are
about to do so (for cone $o$) have higher pressure.  Eq.~\ref{eq:tb}
shows also that when $\Gamma_1\sim\theta_{o}^{-1}$ and the part of the
blast generated by the core of the jet becomes visible to the
observer, cone $o$ has just started spreading.  The contribution of
region $1$ to the light curve of the cone with $\theta_{o}$ is not
dominant because when $\Gamma_1 \sim \theta_{o}^{-1}$ the energy per
unit solid angle is comparable to $\epsilon_{o}$
\begin{equation}
\epsilon_1 \simeq \epsilon_c \left(\frac {\theta_1}{\theta_c}\right)^{-2} 
 \left(\frac {\pi \theta_1^{2}}  {\pi \theta_{o}^{2}}\right)=\epsilon_{o}.
\label{eq:9}
\end{equation} 
The regions with $\theta \gg \theta_{o}$ spread at later time, so when
$\Gamma_3\sim\theta_3^{-1}$ the energy per solid angle of cone $o$ has
already decreased with time but it remains always comparable to
$\epsilon_3$
\begin{equation}
\epsilon_o(t_{b_3})=\epsilon_{o}\,\theta_{o}^{2}\Gamma_o^{2}(t_{b_3})=
\epsilon_{o}\left(
\frac {t_{b_3}}{t_{b_o}}\right)^{-1}= \epsilon_{o} \left(\frac{\theta_3}
{\theta_{o}}\right)^{-2}=\epsilon_3,
\label{eq:dtb}
\end{equation}

\noindent
where $t_{b_i}$ is the break time for cones $i$ and we used
Eq.~\ref{eq:gammatb} and Eq.~\ref{eq:tb}.

Therefore the regions with $\theta \ll \theta_{o}$, (or those with
$\theta$ substantially above $\theta_{o}$), don't determine the
overall shape of the light curve, because they are hidden by the
dominant emission from the component along the line of sight. In
principle, the superposition of many components of energy given by
Eq.~\ref{eq:9} and~\ref{eq:dtb} may give rise to a sizable
contribution. As shown by the more detailed calculation of \S~3.2 (see
Fig~2), this is not the case and the only effect is to delay by a
factor of 2 the time of the break. Consequently the energy per unit
solid angle that is measured modeling the afterglow is $\epsilon_{o}$
and the time break depends only on the viewing angle $\theta_{o}$.

\section{Light curve calculation}

In order to compute a reasonably accurate lightcurve in our model, we
adopt the following approach. We divide the inhomogeneous jet in $N$
hollow cones (all but the very central one), each characterized by
energy per unit solid angle and Lorentz factor given by
Eq.~\ref{eq:top}.  We compute an approximate lightcurve for each
sub-jet using three asymptotic behaviors. For the cones with
$\theta\lsim\theta_o$ and $\theta\gsim\theta_o$ we adopt the
$\theta\ll\theta_o$ and $\theta\gg\theta_o$ approximations described
in \S~2,.  For the cones defined by
$\theta_o-1/\Gamma<\theta<\theta_o+1/\Gamma$ which point towards the
observer since the beginning, we again consider the observer along the
symmetry axis, but the emission is calculated in a filled cone
geometry, instead of the hollow one adopted for the other cases. For
 $\theta>\theta_o-1/\Gamma$ we use the usual afterglow
theory to perform the lightcurve calculations, while for
$\theta\lsim\theta_o$ we generalize it for an off-axis observer as explained
in the following section.

\begin{figure}
\psfig{file=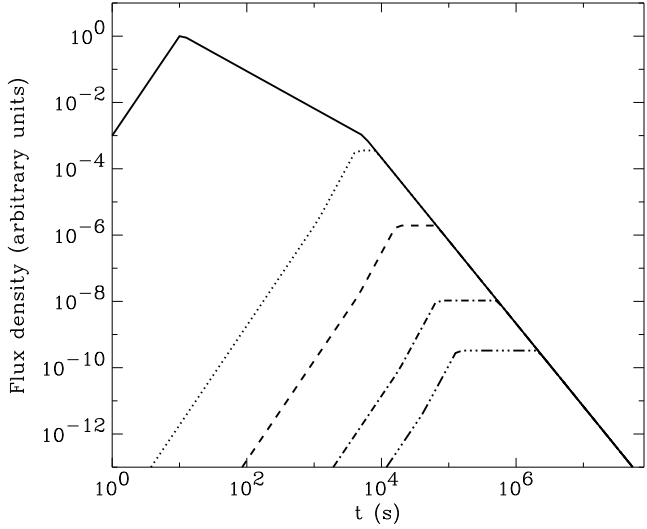 ,width=0.48\textwidth}
\caption{{Light curves of an homogeneous jet (with uniform $\Gamma$ and 
$\epsilon$) as observed from different viewing angles. the geometric
opening angle of the jet is $theta_j=1^\circ$. From top to bottom,
with different line styles, off-axis angles $\theta_o=0$, $2$, $4$ and
$8$ are shown.}
\label{fig:jetto}}
\end{figure}

\subsection{Radiation from an off-axis homogeneous fireball}

\begin{figure}
\psfig{file=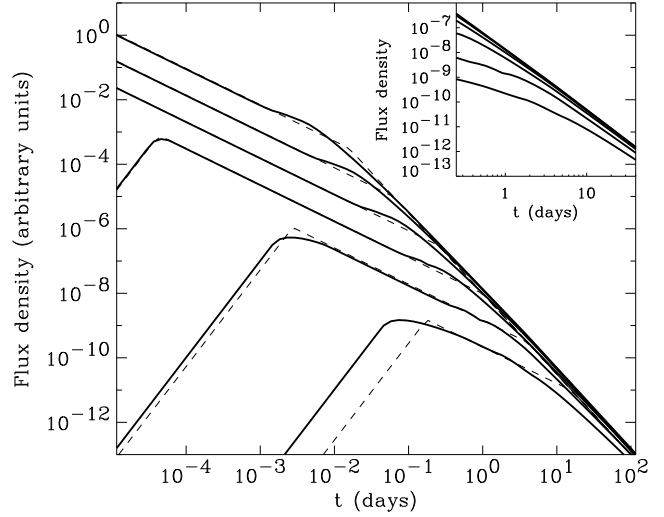,width=0.48\textwidth}
\caption{{The light curves of an inhomogeneous jet observed 
from different angles. From the top we show
$\theta_{o}=0.5,\;1,\;2,\;4,\;8,\;16^{\circ}$.
 The break time is related only to the observer angle:
$t_{b}\propto \theta_{o}^{2}$. The dashed line is the on-axis light
curve of an homogeneous jet with an opening angle $2\theta_{o}$ and an
energy per unit solid angle $\epsilon(\theta_{o})$. The blow up is the
time range between 4 hours and 1 month, where most of the optical
observations are performed.  Comparing the solid and the dashed lines
for a fixed $\theta_{o}$, it is apparent that we can hardly
distinguish the two models by fitting the afterglow data. The slight
flattening of the lightcurves just before the break is likely to be
due to the numerical approximations we adopted (see \S~3).}
\label{fig:jetti}}
\end{figure}

Consider a uniform jet with Lorentz factor $\Gamma$ and initial
half-aperture $\theta_j$. The radiative process is synchrotron emission
(Meszaros \& Rees 1997) and we concentrate on the power law branch of
the spectrum between the peak, $\nu_{m}$, and the cooling, $\nu_{c}$,
frequencies. The observed flux at a frequency $\nu$, time $t$ and
viewing angle $\theta_{o}$ is
\begin{equation}
F( \nu,t,\theta_{o}) \propto A_{e}\,\, I'\left( \frac{\nu}{\delta}, \delta t 
\right)\,\, \delta^{3}( \Gamma,\theta_{o}),
\end{equation}
where $A_{e}$ is the emitting area and
$\delta=(\Gamma\,(1-\beta\cos\theta))^{-1}$ is the relativistic
Doppler factor. $I'$ is the comoving intensity at the comoving frequency
$\nu'=\nu/\delta$ and at the comoving time $t'=\delta t$
\begin{equation}
 I'=I'(\nu'_{m}, t') \Big(\frac{\nu'}{\nu'_{m}}\Big)^{-\alpha} \propto 
\Gamma ^{(2+3\alpha)}\,\, \delta^{(1+\alpha)} \,t \,\nu^{-\alpha}.
\end{equation}
Due to the relativistic beaming, the observed flux depends on the
observer angle. In particular, if $\theta{o}\simeq0$
\begin{equation}
F=F_{on}\simeq \pi\left(\frac{R}{\Gamma}\right)^{2}\,I'\,\,(2\,\Gamma)^{3},
\;\;\;\;\textrm{$\forall$ t}
\label{eq:Fon}
\end{equation}
while for $\theta_{o}>\theta_{j}$
\begin{equation}
F=F_{off}\simeq\left\{\begin{array}{ll}
\pi(R \theta)^{2} I' (\frac{1}{\Gamma (1-\beta\cos\theta_{o})})^{3}& t<t_{b}\\
\pi(\frac{R}{\Gamma})^{2}\,I'\,\,(2\,\Gamma)^{3} &  t > t_{b}.\\	 
                    \end{array} \right.
\label{eq:Foff}
\end{equation}
In this case the jet is initially a source moving at an angle
$\theta_{o}$ with the observer. At the break time $\Gamma\sim
\theta^{-1}$ and then it starts decreasing very fast
(Eq.~\ref{eq:gamma}) until $\theta_{o}-\theta_{j} < \frac {1}
{\Gamma}$ and we can again use the approximation that the jet is
viewed on-axis. So the intensity, the change in the slope and the
break time of the afterglow depend on the observer viewing angle.  In
Fig.~\ref{fig:jetto} we show the resulting off-axis lightcurve for a
jet with opening angle $\theta_j=1^\circ$ as viewed with off-axis
angles $\theta_o=0$, $2$, $4$ and $8$ degrees. The flat part of the
lightcurves corresponds to the time interval between the beginning of
the jet spreading and the time in which the jet enters the line of
sight.  In this time interval our approximations are no longer valid
and we simply model the lightcurve with a flat component connecting
earlier and later times. This crude approximations is likely
responsible for the slight flattening of the final lightcurves (see
below and Fig.~\ref{fig:jetti}) just before the break.
 
\subsection{Inhomogeneous jet}

Using Eq.~\ref{eq:Fon} and Eq.~\ref{eq:Foff} we can compute the total
light curve as a sum over the fireball components according to the scheme
described at the beginning of this section.
\begin{equation}
\begin{array}{l}
L(\nu,t,\theta_{o})=
\sum_{\theta<\theta_{o} - \frac{1}{\Gamma}} \Big[F_{off} 
(\frac{\theta^{2}-\theta_{in}^{2}}{\theta^{2}})\Big]\; + \\ 

\sum_{\theta_{o}-\frac{1}{\Gamma} <\theta<\theta_{o}+\frac{1}{\Gamma}} 
\Big[ F_{on}\,\,\frac{1}{N}\Big] 

+ \sum_{\theta>\theta_{o}+\frac {1}{\Gamma}}\Big[ F_{on}-F_{on}(\theta_{in})\big],
\end{array}
\label{eq:lc}
\end{equation}
where $\theta_{in}$ is the inner edge of the cone and
$F_{on}(\theta_{in})$ is the light curve for a jet with
$\epsilon(\theta)$, $\Gamma(\theta)$ and half-opening angle
$\theta_{in}$. N is the number of jet components that are beamed
around the line of sight at small times.

The results of this calculation are shown in Fig.~\ref{fig:jetti} for
different viewing angles. On top of the solid lines we plot with a
dashed line the lightcurve of a homogeneous fireball with
$\epsilon=\epsilon_o$ and $\theta_j=\theta_o$. This shows that, for
$\theta_o<\theta_j$, the lightcurve of an inhomogeneous jet can be
successfully modelled with the lightcurve expected for a homogeneous
jet: $F\propto t^{-3(p-1)/4}$ for $t<t_b$ and $F\propto t^{-p}$ for
$t>t_b$.

One of the main simplifications that we made in Eq.~\ref{eq:lc} is to
assume that the observer is on the jet axis of the cones with
$\theta\ge\theta_o$.  The main consequence is to predict a transition
between the two power-law branches sharper than what we expect from
the exact integration. This is due to the fact that, even in a uniform
jet, off-axis observers sees smoother transitions (see Fig. 4 of
Ghisellini \& Lazzati 1999).

\section{Time break-energy relation}

Recently an important observational result on the energetic content of
GRB<s was published by PK01 and F01. They found an anti-correlation
between the isotropic equivalent energy, $E_{iso}=4\pi\epsilon$ and
the break time in the afterglow lightcurves.  F01 in particular
derived $E_{iso}$ from gamma ray fluences, $F_{\gamma}\propto E_{iso}$
and their data are consistent with $F_{\gamma}\propto t_{b}^{-1}$.
They explain it in the framework of collimated, uniform, lateral
spreading jets interacting with a constant, low density (0.1
cm$^{-3}$), external medium. The observer is postulated to be along
the jet axis.  They assume that the afterglow emission steepens
because at that time the Lorentz factor has dropped to $\Gamma \sim
\theta_{j}^{-1}$, so that the on-axis observer sees the edge of the
jet and the lateral jet spreading becomes important. In this case
$E_{iso}\propto t^{-1}$ is predicted, which matches observation.  They
convert the observed break times in jet opening angles through the
formulation of Sari et al. (1999).  The $\gamma$-ray energy
measured and corrected for the inferred geometry of the jet is
clustered around $5\times10^{50}$ erg. They concluded that GRBs central
engines release the same amount of energy through jets with very
different opening angles.  In this framework, then, the wide
distributions in kinetic energy per unit solid angle, which spans 3
orders of magnitude, is due only to the distribution of jet solid
angles.  PK01 obtained the same results, modeling a subset of
multiwavelength afterglows from which they could assign an external
density, a time break and an equivalent isotropic energy for the
fireball.  Both F01 and PK01 found geometric angles that span an order
of magnitude but strongly concentrate around $2^{\circ}-4^{\circ}.$ In
the framework of our model an alternative explanation of the observed
relation between $E_{iso}$ and $t_{b}$ can be found.  As discussed in
\S2 and shown in Fig.~\ref{fig:jetti}, we can infer from observations
only the properties of the cone pointing towards the observer and not
those of the whole jet.  So $E_{iso} = 4\pi\epsilon_{o}$ and using
Eq.~\ref{eq:tb} and Eq.~\ref{eq:top} we obtain
\begin{equation}
E_{iso} \propto \theta_{o}^{-2}  \propto  t_{b}^{-1}.
\label{eq:eiso}
\end{equation}
Then, the observed distribution of $E_{iso}$ and its relation with
$t_{b}$ is due to an inhomogeneous jet and the possibility to view it
from different angles.

\section{Luminosity function}

Under our assumptions, each $\gamma$-ray luminosity corresponds to a
particular viewing angle. The probability to see a jet between
$\theta$ and $\theta+d\theta$ is given by
\begin{equation}
P(\theta) d\theta \propto \sin\theta d\theta \;\;\;\;\; 0 \leq \theta \leq \theta_{j},
\label{eq:p}
\end{equation}
$\langle\theta\rangle\simeq 0.7$ and the highest probability is for
$\theta=\theta_{j}$.  Therefore it is highly improbable to see a jet
on axis. Consequently we expect more faint GRBs then very luminous
ones according to a luminosity function
\begin{equation}
P(y)\propto 10^{-y},
\end{equation}
where $y=\log\epsilon$.

Since there are only a small number of GRBs with observed red-shift,
the comparison of our predicted luminosity function with data is far
from being definitive.  Recently Bloom at al. (2001) published an
histogram of the bolometric k-corrected prompt energies for 17 GRBs.
The distribution is roughly flat from $6 \times 10^{51}$  to $2
\times 10^{54}$ erg but as the authors emphasize this analysis applies
only to observed GRBs with redshifts and several observational biases
obscure the true underlying energy distribution. The main bias that
overcasts faint GRBs is the detection threshold of the instruments:
this sample is thus flux and not volume limited. Moreover redshift
determination encounters more problem for faint GRBs.  Based on
$\langle V/V_{max} \rangle$-hardness correlation, Schmidt (2001)
derived a luminosity function without using any redshift. They had to
assume how the comoving GRBs rate varies with redshift and they based
their calculations on three star forming rate models.  In this case a
power-law luminosity function was derived, but flatter than the
$n(L)\propto L^{-2}$ predicted in the simplest version of our model.
A luminosity function not based on assumed burst rate evolution can be
derived by measuring the burst distance scale through the recently
discovered variability-luminosity relation (Fenimore \& Ramirez-Ruiz
2000, Reichart et al. 2001). A cumulative analysis of a sample of 220
bursts (Fenimore \& Ramirez-Ruiz 2000) yielded a power-law luminosity
function $n(L)\propto L^{-2.33}$, which compare more favorably with
our prediction. More recently, Lloyd-Ronning, Fryer \& Ramirez-Ruiz
(2001) find that the typical luminosity of GRBs evolves with
redshift. As a consequence, a flatter power-law index $n(L)\propto
L^{-2.2}$ is obtained.  We stress, however, that our deduced
luminosity function can be altered by distributions in total energy or
geometric angles or by luminosity evolution with redshift.  Given the
somewhat contradictory observational results discussed above, we
conclude that more accurate spectral and fluence measures and a larger
sample of bursts are needed for a proper comparison.

\section{Discussion and conclusions}

We considered inhomogeneous GRBs jets with a standard total energy,
opening angle and local energy distribution, $\epsilon\propto\theta ^{-2}$.
We show that this jet structure can reproduce the observed correlation
between isotropic energy and break-time.  In this model both
measurements depend only on the viewing angle because the $\gamma$-ray
fluence and the afterglow emission are dominated by the components of
the jet pointing towards the observer at small times. Since all cones
have the same total energy $E_{\rm jet}=2\pi\theta^{2}\epsilon\,=cost$, we
recover the results of F01 and PK01 but the constrains on the
geometrical beaming can be relaxed and an appealing more standard
structure for all GRBs can be adopted. The jet total energy can be
calculated from Eq.~\ref{eq:top}
\begin{equation}
E_{Total}=2\pi\epsilon_{c} \theta_{c}^{2}\,\,\left(1+2 
\ln\frac{\theta_{j}}{\theta_{c}}\right) =
E_{\rm jet} \,\left(1+2 
\ln\frac{\theta_{j}}{\theta_{c}}\right)
\end{equation}
and compared to $E_{\rm jet}=2\pi\theta_{o}\epsilon_{o}=2\pi\epsilon_{c}
\theta_{c}^{2}$, the total energy inferred from observation, $E_{\rm jet}
\leq E_{Total}$. To give an example, for a fireball with
$\theta_c=1^\circ$ and $\theta_j=20^\circ$, we have $E_{Total}/E_{\rm
jet}\simeq6$, i.e. the true energy of the fireball can be one order of
magnitude larger than what inferred with the models of F01 and PK01.

In addition (\S~5) we can derive the GRBs luminosity function from the
probability distribution of the viewing angle and compare it to
data. The comparison is, at this stage, still uncertain because more
accurate spectral and fluence measures are required to build a volume
limited sample and confirm (or rule out) this model.

In any fireball model considering a jet-like GRB structure, a
certain number of afterglows without $\gamma$-ray emission (orphan
afterglows) is expected.  For an homogeneous jet, orphan afterglows
are possible only for viewing angles greater than $\theta_j+1/\Gamma$.
In our jet configuration a fraction of the total area could have a
Lorentz factor lower than the minimum $\Gamma$ necessary for
$\gamma$-ray radiation.  Consequently, considering the same opening
angle, an inhomogeneous jet could produce an higher fraction of orphan
afterglows than an homogeneous one.  This intrinsic fraction depends
above all on the $\Gamma$ distribution within the jet
($\alpha_{\Gamma}$ in Eq~\ref{eq:gammadis}) and on the minimum Lorentz
factors to produce $\gamma$-ray radiation and afterglow emission.  The
observed number of orphan afterglows depends also on flux detection
limits, GRB explosion rates with redshift and cosmology. An accurate
calculation of the expected orphan afterglow rates is therefore beyond
the scope of this paper.

We emphasize that we implicitly assumed the radiation efficiency of the
fireball to be weakly dependent on $\Gamma$ or $\epsilon$. This is a
plausible assumption in internal-shocks scenario. If it was not the
case and the efficiency grew with $\epsilon$, a different relation
between $\epsilon$ and $\theta$ should be postulated in order to
reproduced observations.
\begin{figure}
\psfig{file=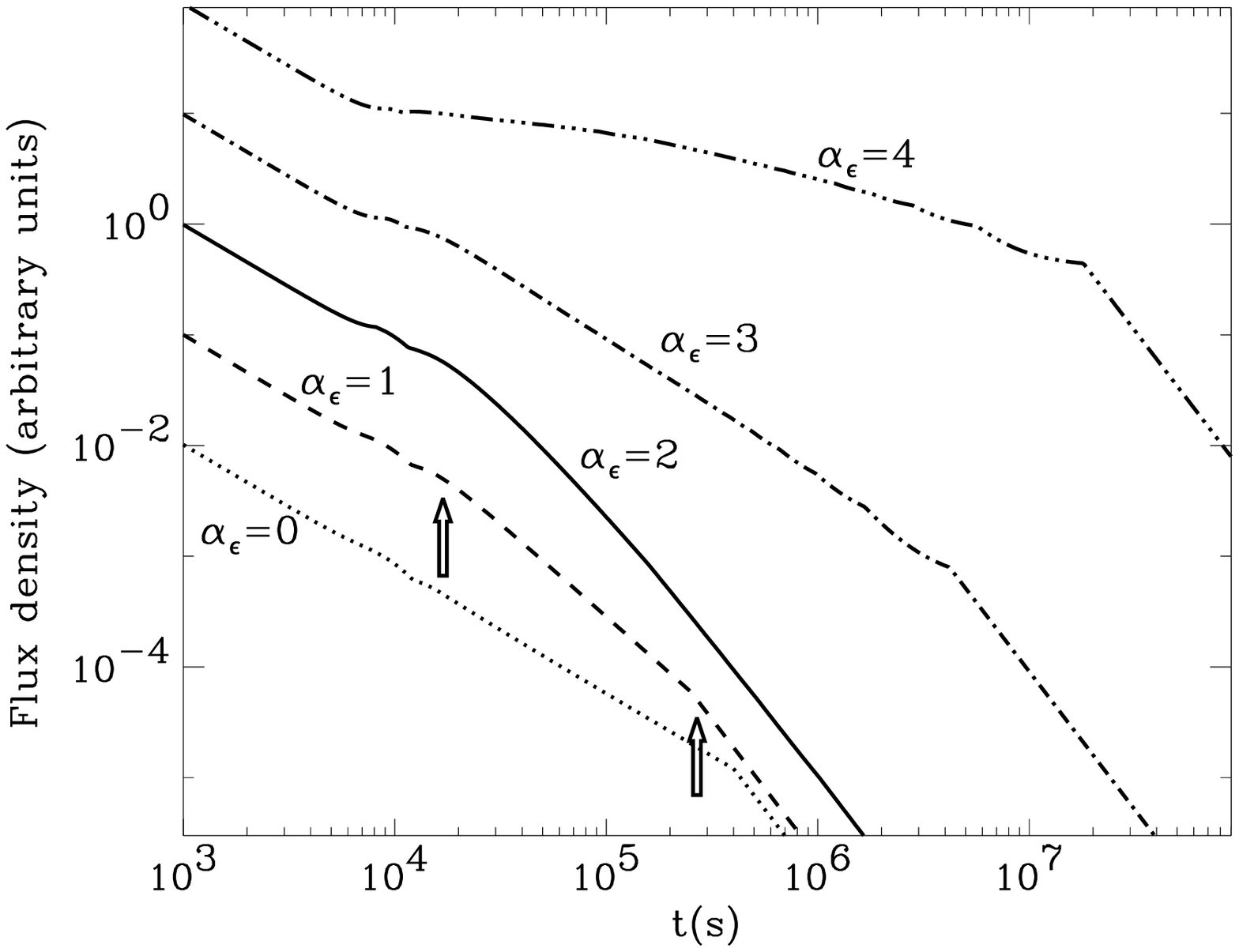,width=0.48\textwidth}
\caption{{The lightcurves of inhomogeneous jets with different indices
 $\alpha_\epsilon$ ($\epsilon\propto\theta^{-\alpha_\epsilon})$.
 The lightcurves have the same $\epsilon_o$ but they are plotted shifted
by factors of 10 for clarity (they would be indistinguishable at small time).
The arrows highlight the location of the two breaks for $0<\alpha<2$. }
\label{fig:alfae}}
\psfig{file=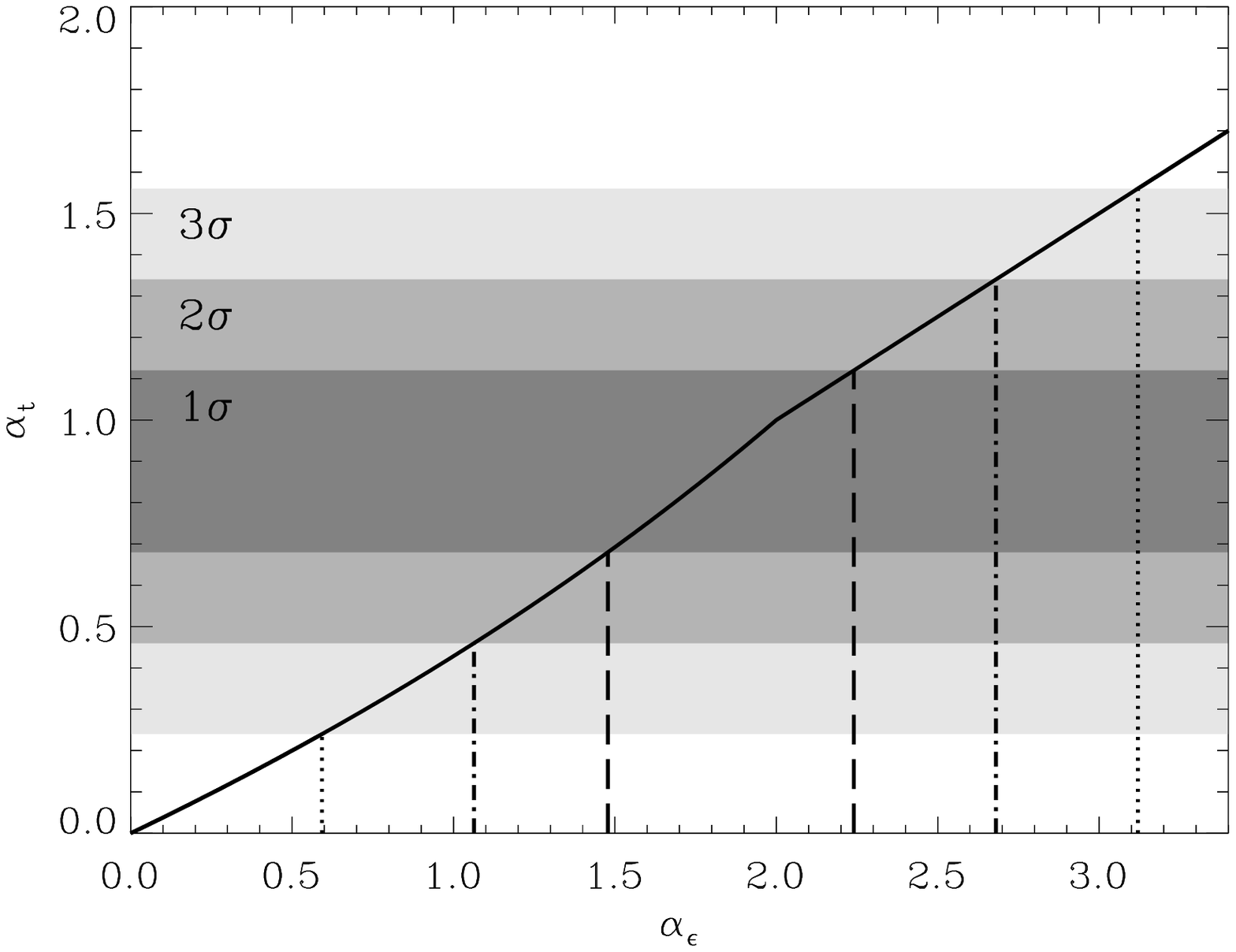,width=0.48\textwidth}
\caption{{The relation between the indices $\alpha_t$ ($E_{iso}\propto
t_b^{-\alpha_t}$) and $\alpha_\epsilon$
($\epsilon\propto\theta^{-\alpha_\epsilon}$) from an inhomogeneous
jet. The shaded regions show the best fit to F01 data
$\alpha_t=0.9\pm0.22$.}
\label{fig:indi}}
\end{figure}
In this paper we concentrated for simplicity on a beam profile
$\epsilon\propto\theta^{-2}$ which is consistent observational results
but it is interesting to briefly discuss other power-law relations
$\epsilon\propto\theta^{-\alpha_\epsilon}$ (see Fig.~\ref{fig:alfae}).
 A decay flatter then 2
would cause two breaks in the light curve: the first due to the cone
pointing the observer when $\Gamma(\theta_{o})\sim\theta_{o}^{-1}$ and
the second, at later times, when the observer sees the edge of the jet
and $\Gamma(\theta_{j}) \sim \theta_{j}^{-1}$. The power law index after
the first break is flatter then $t^{-p}$ because the cones with
$\theta_{o}<\theta\leq\theta_{j}$ enter the line of sight with
$\epsilon \geq \epsilon_{o}$ and substantially modify the light curve
shape.  With a steeper decay in the distribution of $\epsilon$ the
time break and the emission after that break would be dominated by the
jet along the axis rather then by the very much weaker part directed
to the observer. The jet break would then be preceded by a prominent
flattening in the lightcurve, especially 
for $\alpha_\epsilon>3$, difficult to reconcile with observations.
 In this case we would have $\gamma$-ray emission only for very
small angles and the number of orphan afterglows would be much greater
then expected from the $\alpha_\epsilon=2$ model.
From Eq.~\ref{eq:tb} we can derive the
relation between the index $\alpha_\epsilon$ and $\alpha_t$ (where
$E_{iso}\propto t_b^{\alpha_t}$). We obtain
$\alpha_t=\alpha_\epsilon/2$ for $\alpha_\epsilon\ge2$ and
$\alpha_t=3\,\alpha_\epsilon/(8-\alpha_\epsilon)$ for
$\alpha_\epsilon<2$, when the first break is considered. In
Fig.~\ref{fig:indi} we plot this relation overlaid on the interval in
$\alpha_t$ allowed by observations (we used F01 data). We derive
$1.5\lsim\alpha_\epsilon\lsim2.2$, at the $1\sigma$ level. Further
$\gamma$-ray and afterglow observations will allow to constrain this
parameter much better in the future.

Besides the luminosity function discussed in \S~5, there are several
ways in which this model can be proved or disproved.  First, we have
shown that the real total energy of the fireball can easily be an
order of magnitude larger than what estimated by PK01 and F01. In this
case, after the fireball has slowed down to mild-relativistic
and sub-relativistic speed, radio calorimetry (Frail, Waxman \&
Kulkarni 2000) should allow us to detect the excess energy. In
addition, in this model we naturally predict that the more luminous
part of the fireball have higher Lorentz factors. This may help
explaining the detected luminosity-variability (Fenimore \&
Ramirez-Ruiz 2000) and luminosity-lag (Norris, Marani \& Bonnell
2000) correlations ( Salmonson 2000, Kobayashy, Ryde \&
MacFadyen 2001; Ramirez-Ruiz
\& Lloyd-Ronning 2002). Another constrain is given by
polarization. Since fireball anisotropy is a basic ingredient of this
model, inducing polarization (Ghisellini \& Lazzati 1999; see
also Sari 1999 and Gruzinov \& Waxman 1999).  The time evolution of
the polarized fraction and of the position angle are however different
from a uniform jet (Rossi et al. in preparation). Finally, the
properties of the bursts should not depend on the location of the
progenitor in the host galaxy, and therefore this model can
accommodate the marginal detection of an $E_{\rm iso}$-offset relation
(Ramirez-Ruiz, Lazzati \& Blain 2001) only if a distribution of
$\theta_j$ is considered.

\section*{Acknowledgments}
We thank G. Ghisellini, J. Granot, P. Kumar, A. Panaitescu and
E. Ramirez-Ruiz for many stimulating discussions. ER and DL thank the
IAS, Princeton, for the kind hospitality during the final part of the
preparation of this work. ER thanks the Isaac Newton and PPARC for
financial support.

\end{document}